\documentclass[letterpaper, journal, twoside]{IEEEtran}

\IEEEoverridecommandlockouts

\usepackage{dsfont}
\usepackage[T1]{fontenc}
\usepackage{graphicx, stfloats} 
\usepackage{svg}
\usepackage{geometry}
\usepackage{standalone}
\usepackage{hyperref}
\usepackage{pgfplots}
\usepackage{booktabs}
\usepackage{subcaption}
\usepackage{amssymb, amsmath}
\usepackage{algorithm}
\usepackage{algpseudocode}
\usepackage{makecell}
\usepackage{bbm}
\usepackage{diagbox, multirow} 
\usepackage{color,soul} 
\usepackage{siunitx}
\usepackage{xurl}

\newcommand{\revision}[1]{\textcolor{black}{ #1}}

\definecolor{vandeusen}{RGB}{73,92,111}
\definecolor{cordovan}{RGB}{152,68,71}
\definecolor{alizarin}{rgb}{0.82, 0.1, 0.26}
\definecolor{azure}{rgb}{0.0, 0.5, 1.0}

\title{Morphing Wing Designs in Commercial Aviation}

\author{Chengyue Dong$^1$, Mansur M. Arief$^2$%

\thanks{$^1$Chengyue Dong is with the Saratoga High School, Saratoga, CA, USA}%
\thanks{$^2$Mansur M. Arief is with the Department of Aeronautics and Astronautics, Stanford University, Stanford, CA, USA (\url{mansur.arief@stanford.edu})
}%
\thanks{$^{\star}$Corresponding Author}%
}

\begin{document}

\markboth{Preprint Version. January, 2025}
{Dong \& Arief (2025): Morphing Wing Designs in Commercial Aviation}

\maketitle

\begin{abstract}
\revision{With increasing demands for fuel efficiency and operational adaptability in commercial aviation}, this paper provides a systematic review and classification of morphing wing technologies, analyzing their aerodynamic performance characteristics and atmospheric condition adaptability. We first develop a comprehensive classification framework for morphing wing designs based on their scale of morphing, actuation mechanisms, and intended purposes. Through analysis of historical developments and current implementations, we evaluate two significant case studies: the Mission Adaptive Compliant Wing (MACW) and Adaptive Aspect Ratio (AdAR) morphing wing, demonstrating performance improvements of up to 25\% in drag reduction and 40\% in control authority. \revision{Our investigation reveals critical trade-offs between full-span and partial morphing approaches, particularly regarding implementation complexity, certification requirements, and operational reliability.} The study concludes with an assessment of technical barriers and opportunities, providing specific recommendations for advancing morphing wing technology in commercial aviation applications. \revision{Key findings indicate that while material science and control system advances enable practical implementation, certification pathways and maintenance considerations remain critical challenges for widespread adoption.}
\end{abstract}


\section{Introduction}

\IEEEPARstart{T}{he} evolution of aviation has been defined as a relentless pursuit of innovation, endeavoring for faster, more efficient, and more adaptable aircraft with the highest safety level. Air travel demand is expected to continue to increase in the next 20 years at a rapid pace \cite{boeing2023cmo}, and engineers and researchers are constantly exploring new innovative technologies and design concepts to meet these challenges, including designing by replacing about half of the global fleet with new, more fuel-efficient models \cite{investors2023boeing}. To this end, morphing wing design is one of these promising new innovative designs, which allows the wings of an aircraft to change their shape and size in response to different flight conditions and environments. 

Traditionally, aircraft wings are designed with fixed sizes and shapes, optimized for certain conditions, but leave drawbacks for other conditions. Taking a simple example, the wings of modern airliners are swept moderately toward the rear of the aircraft, which are designed for commercial long hauls that optimize the high-altitude cruising stage of flight. These large aircraft will have difficulty being efficient in all stages of flight compared to private aircraft \cite{gorn2021high}. In contrast, fighter jet wings are swept much farther back to enable maneuverability and supersonic travel but will suffer flying at low speeds and altitudes \cite{gorn2021high}. 

\begin{figure}[t!]
    \centering
    \includegraphics[width=\linewidth]{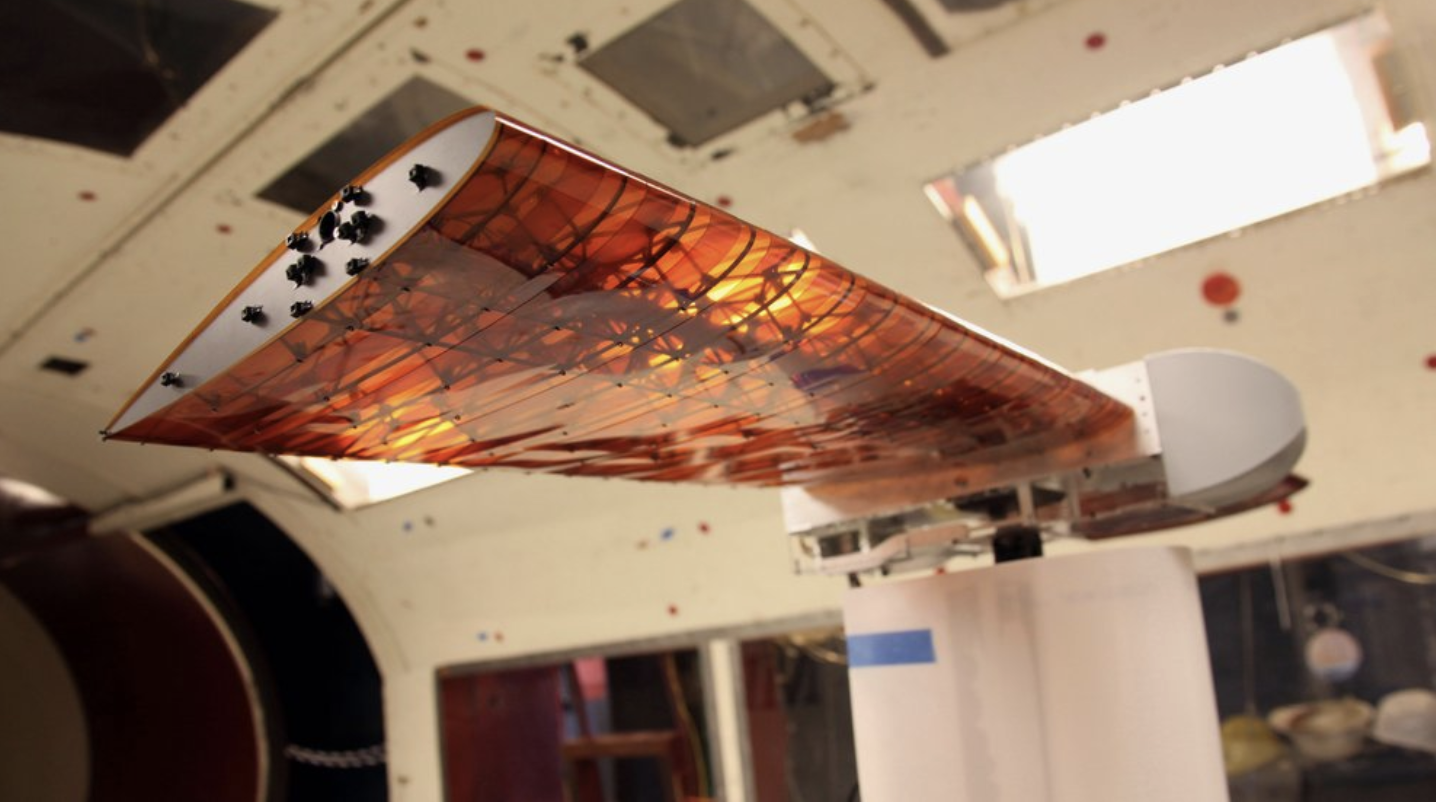}
    \caption{An example of an aircraft morphing wing design~\cite{choi2017idea}}
    \label{fig:morphing_wing_design_example}
\end{figure}

\revision{The commercial aviation sector is currently experiencing} a shift towards a wider range of aircraft functionalities beyond traditional long-haul passenger flights. Industry research companies predict increasing demand for regional jets and turboprop aircraft in their market forecasts \cite{frost2022global}, while major aircraft manufacturers like Boeing have an increasing focus on developing versatile aircraft that can be configured according to load \cite{boeing2015c40}. These trends indicate a shift in the commercial aviation industry towards adaptability and diversification, aiming for a morphing sign design that allows airplanes to do all types of missions \cite{milstein2007shape} and cater to the evolving needs of the market. 

In addition to the evolving market, the aviation industry is also facing increasing challenges to reduce its environmental impact and operational costs while maintaining aircraft safety and reliability \cite{merlin2020green}. A significant factor that affects the performance of an aircraft in a certain condition is the wing design of the aircraft, which largely determines the aerodynamic characteristic of the aircraft. \revision{Current fixed-wing designs have reached their aerodynamic limits}, and thus will not be able to meet the rapidly growing demand for more fuel-efficient aircraft among other demands. To this end, fixed-wing aircraft have limited capabilities to cope with varying wind speeds, turbulence, and gusts, which can compromise their stability and safety. Furthermore, fixed-wing aircraft may be less suitable for some applications, such as unmanned aerial vehicles (UAVs) \cite{ling2019aerial}, electric aircraft, and low-speed flight, where flexibility and versatility are required. In contrast, a morphing or shape-changing wing has the potential to be optimal across all flight conditions, thus reducing fuel consumption \cite{choi2017idea}. In Fig.~\ref{fig:morphing_wing_design_example}, an example of morphing wing design is given. With this type of technology, one can optimize the lift-to-drag ratio for various angles of attack, giving an increased performance of up to 18.3\% compared to conventional wing designs \cite{ninian2017design}.

\revision{This paper provides a comprehensive review of morphing wing designs in modern aircraft for domestic applications, with particular emphasis on their advantages in dealing with variable environments. The remainder of this paper is structured as follows.} Section~\ref{sec:related_work} \revision{presents} a literature review of morphing wing design with a focus on the limitations of domestic application, and why it is important, including its history, existing technologies, and application in aerospace. Section~\ref{sec:framework} \revision{describes} a theoretical framework of morphing technology, including its definition, classification, types, and principles. Section~\ref{sec:case_studies} \revision{examines} the case studies of existing aircraft that incorporate morphing wing technology, and Section~\ref{sec:discussion} highlights their implementations and technical challenges. \revision{Finally}, Section~\ref{sec:conclusion} concludes and summarizes key findings, implications, and recommendations for future research and development.

\section{Literature Review}\label{sec:related_work}

\revision{The development of morphing wing technology represents a century-long evolution in aerospace engineering, from primitive mechanical solutions to sophisticated adaptive systems. This review traces the progression of wing morphing concepts, examining how each advancement has contributed to our current understanding and highlighting the technological challenges that remain to be addressed.}

\begin{figure}
    \centering
    \includegraphics[width=\linewidth]{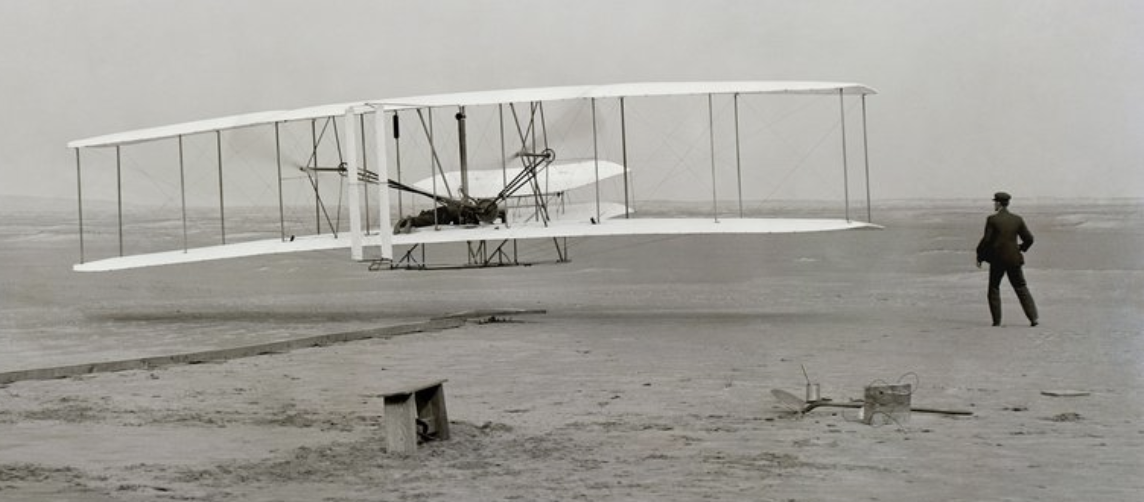}
    \caption{Wright’s brothers primitive morphing wings ~\cite{choi2017idea}}
    \label{fig:wrights_morphing_wing_design}
\end{figure}

\subsection{Early Development of Wing Morphing}
While the idea of a morphing wing design may seem futuristic, the method of controlling an aircraft's flight movements by changing the shape of the wings has a surprisingly long history. This pursuit can be traced back to the very beginning of aircraft, with the Wright Brothers playing a key role in the development of this concept \cite{choi2017idea, nasm2022researching}. Fig.~\ref{fig:wrights_morphing_wing_design} shows an early example of the implementation of this concept. 

\begin{figure}[h!]
    \centering
    \includegraphics[width=0.7\linewidth]{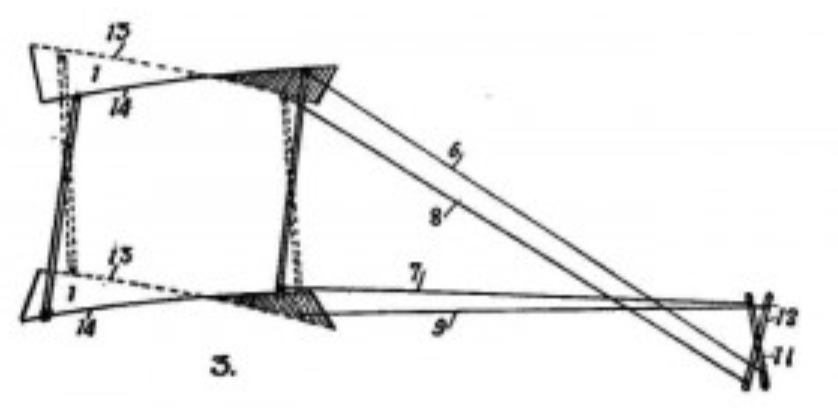}
    \caption{The Wright brother’s warping wing sketch~\cite{biolocomotion2011wright}}
    \label{fig:wrights_morphing_wing_sketch}
\end{figure}

\revision{The Wright brothers' pioneering work emerged from} a fundamental challenge of how to control aircraft during their early testing on kites and small gliders \cite{nasm2022researching}. Their experiments revealed that air currents were too strong for human reflection, and the aircraft must be inherently stable and controllable by humans \cite{nasm2022researching}. \revision{This led to the development of} wing-warping, a method they first tested on their 1899 kite (see Fig.~\ref{fig:wrights_morphing_wing_sketch}).

\subsection{Technical Evolution of Wing Control}
\revision{The wing-warping mechanism fundamentally relies on} spatial modeling of the aircraft structure. The design enables controlled movement of each wing's trailing edge through a system of wires connecting the wingtips to the pilot's footrest. When the pilot operates the pedals, the wires adjust the wing tip position, altering the outer panel's shape. This change in the airfoil's deflection angle creates differential lift forces between the wings, causing the aircraft to rotate about its center of gravity \cite{nasa2023warping}. This innovative design concept, implemented in the Flyer, achieved the first powered flight in 1903.

\subsection{Emergence of Aileron Technology}
The development of control surfaces for roll management, particularly the aileron, evolved through various iterations in the late 19th and early 20th centuries \cite{crouch2008oldies}. \revision{The progression began with} Matthew Piers Watt Boulton's patent in the 1860s, followed by contributions from Charles Renard and Alphonse Pénaud \cite{crouch2008oldies}. Jean-Marie LeBris's experiments with wing warping further advanced the field, ultimately leading to the Wright brothers' breakthrough design.

\revision{A significant development came from} French experimenter Robert Esnault-Pelterie, who introduced separate ailerons positioned between the wings, prioritizing simplicity and reducing structural strain. The term "aileron," meaning "little wing" in French, wasn't adopted until 1908 \cite{crouch2008oldies}. \revision{This innovation represented a crucial shift from wing-warping to discrete control surfaces, offering improved reliability and maintainability.}

\subsection{Modern Variable Geometry Wings}
\revision{The F-14 Tomcat represents a significant milestone in variable geometry wing design}. Its wings could pivot between 20$^{\circ}$ during takeoffs and landings to a maximum of 68$^{\circ}$ sweep angle for reduced drag at supersonic speeds \cite{airforce2020f14}. This design provided exceptional versatility, enabling high-speed performance and improved maneuverability across various flight conditions.

The wing transformation system utilized the Standard Central Air Data Computer (SCAD), which calculated appropriate sweep angles based on altitude and Mach number. \revision{The system's sophisticated control mechanism} achieved precise wing adjustments through a hydro-mechanical system operating at eight degrees per second \cite{airforce2020f14}. \revision{While this design demonstrated remarkable capabilities, it also highlighted the challenges of complex morphing systems, including increased maintenance requirements and reduced fuel storage capacity.}

\subsection{Recent Developments}
A significant advancement occurred on November 15th, 2002, when NASA tested their adaptation of wing-warping control on an F-18, designated as the X-53. This Active Aeroelastic Wing program focused on developing flexible wing warping for roll control \cite{nasa2023aeroelastic, dibley2005development}. The Air Force confirmed the program's success, noting its potential applications in both military and high-altitude long-endurance aircraft \cite{barr2005wing}.

\revision{Contemporary research has produced innovations such as} the Adaptive Aspect Ratio morphing wing (AdAR) at Swansea University \cite{woods2015adaptive}. \revision{This design represents a modern approach to shape-shifting wing technology, combining advanced materials and control systems to enhance aircraft performance and efficiency. The AdAR concept demonstrates the continuing evolution of morphing wing technology, suggesting promising directions for future aerospace applications.}

\revision{Despite these significant advances in morphing wing technology, several critical challenges persist in implementing these systems in commercial aviation. While military applications have demonstrated the potential benefits of variable geometry wings, the commercial sector faces unique constraints related to maintenance costs, reliability requirements, and safety certifications. Furthermore, the increasing demand for fuel efficiency and operational flexibility in commercial aviation creates an urgent need for innovative solutions. This study addresses these challenges by examining current morphing wing technologies and their potential applications in commercial aircraft, with particular attention to practical implementation considerations and economic viability. Understanding these aspects is crucial for advancing the field and developing solutions that can meet the demanding requirements of commercial aviation operations.}

\section{Framework and Designs}\label{sec:framework}

\begin{table*}[h!]
\centering
\caption{Morphing wing design summary}
\label{tbl:wing_design_summary}
\begin{tabular}{@{}lll@{}}
\toprule
\multicolumn{1}{c}{\multirow{2}{*}{$\quad$$\quad$$\quad$$\quad$$\quad$\textbf{Criteria}$\quad$$\quad$$\quad$$\quad$$\quad$}} & \multicolumn{2}{c}{\textbf{Wing Designs}}                                                 \\ \cmidrule(l){2-3} 
\multicolumn{1}{c}{}                                   & \multicolumn{1}{c}{$\quad$$\quad$$\quad$$\quad$$\quad$$\quad$\textbf{Full-Span Morphing}}$\quad$$\quad$$\quad$$\quad$$\quad$$\quad$ & \multicolumn{1}{c}{$\quad$$\quad$$\quad$$\quad$$\quad$$\quad$\textbf{Partial Morphing}$\quad$$\quad$$\quad$$\quad$$\quad$$\quad$}             \\ \midrule
Shape Memory Alloys                                    & Open research area                     & Potential use in SMA-based control surfaces      \\
Flexible Structures                                    & Adaptive Aspect Ratio (AdAR) morphing wing            & Mission Adaptive Compliant Wing (MACW)                     \\
& (Complete adaptability) &  (Localized adaptability) \\
Purpose                                                & Efficiency enhancement                 & - Performance optimization\\ 
& & - Mission adaptability \\ \bottomrule
\end{tabular}
\end{table*}

\revision{Morphing wing technology represents a diverse field encompassing various design approaches and implementation strategies. To systematically analyze these approaches, we propose a classification framework based on morphing scale, mechanism type, and intended purpose. This framework enables better understanding of design trade-offs and application-specific requirements.}

\subsection{Classification by Scale of Morphing}
\revision{Morphing wing designs can be categorized along a continuum from comprehensive to localized transformations.} As shown in Table~\ref{tbl:wing_design_summary}, full-span morphing represents one end of this spectrum, where wings are capable of dynamically adjusting their entire geometry. The Adaptive Aspect Ratio (AdAR) morphing wing \cite{woods2015adaptive} exemplifies this approach, offering complete adaptability across the wing structure. At the other end of the spectrum are partial morphing designs, such as the Mission Adaptive Compliant Wing (MACW) \cite{flexsys2024flexfoil}, which incorporate more localized morphing mechanisms that selectively alter specific wing sections to achieve desired aerodynamic effects.

\subsection{Morphing Mechanisms}
The implementation of morphing capabilities employs diverse technologies for actuation and shape alteration. Shape Memory Alloys (SMA) represent an emerging technological approach, where temperature changes trigger wing morphology alterations \cite{afshohteh2023sma}. While full-span SMA implementation remains an open research area, these materials show immediate promise in partial morphing applications, particularly for control surfaces. Flexible structures offer an alternative approach, achieving dynamic shape changes in response to aerodynamic forces. This technology has been successfully demonstrated in both full-span applications through the AdAR system and partial implementations via the MACW design.

\subsection{Purpose-Based Classification}
\revision{The intended purpose of morphing wing designs provides crucial context for their practical implementation in commercial aviation.} As illustrated in Table~\ref{tbl:wing_design_summary}, full-span morphing designs primarily target efficiency enhancement through comprehensive wing adaptation. These systems optimize aerodynamic performance across various flight conditions by enabling complete wing reconfiguration. In contrast, partial morphing approaches serve dual purposes: performance optimization through localized adaptations and mission adaptability through selective control surface modification. This distinction in purpose reflects the different operational requirements and design constraints faced by commercial aircraft.

The classification framework presented in Table~\ref{tbl:wing_design_summary} demonstrates the interrelation between morphing scale, mechanism choice, and intended purpose. \revision{Full-span morphing, while offering the greatest potential for efficiency gains, requires more complex implementation strategies and faces greater certification challenges. Partial morphing approaches, though more limited in scope, provide practical advantages in terms of implementation feasibility and certification pathways.} This categorization aligns with the objectives outlined by \cite{friswell2006morphing} and reflects the primary motivations driving morphing wing development in commercial aviation.

\revision{The diverse nature of these design approaches highlights the complexity of implementing morphing wing technology in commercial aviation. While each approach offers distinct advantages, the selection of appropriate morphing strategies must consider factors such as reliability, maintainability, and certification requirements. Understanding these trade-offs between full-span and partial morphing approaches, along with their associated mechanisms and purposes, provides a foundation for developing practical morphing wing solutions that meet the demanding requirements of commercial aircraft operations.}

\section{Case Studies}\label{sec:case_studies}

\revision{This section examines two significant implementations of morphing wing technology that demonstrate different approaches to achieving wing adaptability: the Mission Adaptive Compliant Wing (MACW) and the Adaptive Aspect Ratio (AdAR) morphing wing. These cases illustrate the practical challenges and opportunities in morphing wing design.}

\subsection{Mission Adaptive Compliant Wing}
The Mission Adaptive Compliant Wing (MACW), developed by FlexSys and tested by the Air Force Research Laboratory, demonstrates the potential of elastic materials in achieving wing shape modification during flight \cite{flexsys2024flexfoil}. The system's seamless, hinge-free design enables real-time adaptation of trailing and leading edges to variable flight conditions. Founded by Sridhar Kota at the University of Michigan, FlexSys developed this approach to exploit the elasticity of aviation-grade materials through distributed compliance \cite{flexsys2024flexfoil}. This design philosophy emerged from the observation that modern commercial airlines allocate approximately 25\% of their operating expenses to fuel costs, while conventional wing and engine designs approach their efficiency limits \cite{kota2024future}.

The MACW's technical innovation lies in its compliant structure mechanism, which represents a significant departure from traditional designs. Unlike previous approaches such as the F-111 Aardvark that relied on rigid link mechanisms, the MACW achieves wing deformation as a unified structure. This design eliminates the complexities and stress concentrations associated with traditional hinged designs \cite{smith1992variable}. The system integrates actuators and sensors within a unified material design, enabling large deformations from small strains.

Flight testing with NASA's Gulfstream III has demonstrated the system's capabilities, achieving a 40\% increase in control authority per degree deflection and up to 25\% reduction in drag. The system successfully operated under maximum dynamic pressure of 1,875 kg/m$^2$. The Air Force Research Laboratory's evaluation confirmed MACW's compatibility with long-endurance aircraft operations in high-altitude, low-temperature environments \cite{kota2009mission}. These results support NASA projections of potential industry-wide fuel savings of up to \$250 billion through such "green" aviation technologies \cite{merlin2020green}.

\subsection{Adaptive Aspect Ratio Morphing Wing}
The Adaptive Aspect Ratio (AdAR) morphing wing, developed at Swansea University, presents an alternative approach through its innovative telescopic spar structure \cite{woods2015adaptive}. \revision{Table~\ref{tbl:adar_characteristics} summarizes the key design elements and performance characteristics of the AdAR system.}

\begin{table}[t]
\caption{AdAR System Characteristics and Performance Metrics}
\label{tbl:adar_characteristics}
\begin{tabular}{p{0.45\linewidth}p{0.45\linewidth}}
\toprule
\textbf{Design Element} & \textbf{Characteristics} \\
\midrule
Structural Design & Telescopic spar with discontinuous geometry; Sliding ribs with distributed support network \\
\midrule
Actuation System & Tension-driven strap drive; High-strength fabric straps \\
\midrule
Material Innovation & Elastomeric matrix composites; >150\% recoverable strain; Transverse fiber reinforcement \\
\midrule
Performance Features & Minimal actuation requirements; Enhanced structural integrity; Efficient load distribution \\
\bottomrule
\end{tabular}
\end{table}

\begin{figure}
    \centering
    \includegraphics[width=\linewidth]{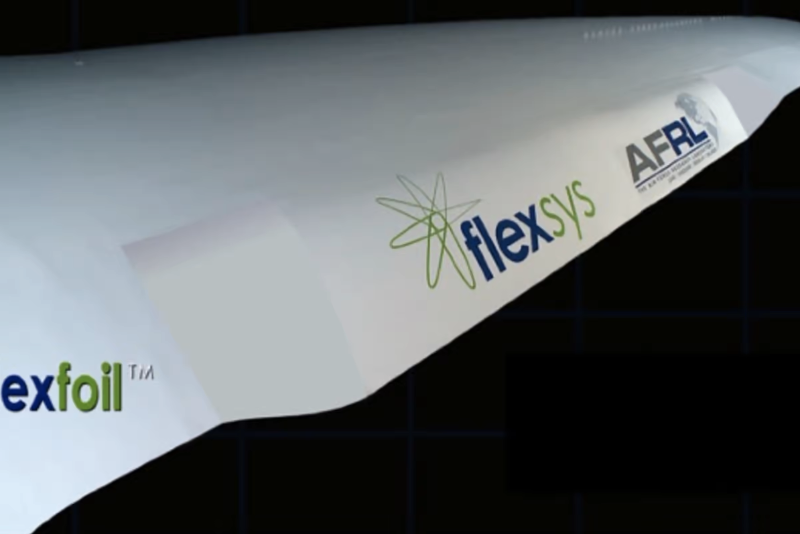}
    \caption{FlexSys morphing wing design~\cite{flexsys2024flexfoil}}
    \label{fig:flexsys_morphing_wing}
\end{figure}

The AdAR system's significance lies in its potential impact on High Altitude Long Endurance Aircraft operations. While modern commercial aircraft achieve optimal aerodynamic characteristics at specific flight conditions, the AdAR system enables adaptation across varying flight phases and fuel conditions. This adaptability proves particularly valuable for aircraft with high fuel weight proportions, where performance optimization throughout the flight envelope becomes crucial \cite{woods2015adaptive}.

\revision{The implications of these morphing wing technologies extend beyond traditional aviation applications. For instance, in unmanned aerial vehicle operations, particularly medical delivery services, the enhanced efficiency and adaptability offered by morphing wings could significantly improve service range and reliability. Studies indicate that such improvements could enhance healthcare access in remote areas through more efficient drone operations} \cite{ling2019aerial}.

Both case studies highlight critical considerations for commercial implementation, including material fatigue, position repeatability, maintenance requirements, and certification compliance. The MACW's successful flight testing and the AdAR's innovative design approach demonstrate the feasibility of different morphing wing solutions, while also emphasizing the importance of addressing long-term reliability and operational requirements in commercial aviation applications.

\section{Discussion}\label{sec:discussion}
\revision{The implementation of morphing wing technology in commercial aviation faces three primary technical challenges. First, material durability and fatigue resistance must meet aviation standards while maintaining morphing capabilities. Second, control system integration must ensure precise and reliable shape adaptation across varying flight conditions. Third, certification pathways must be established for these novel structural approaches.}

The MACW and AdAR implementations demonstrate different approaches to addressing these challenges. The MACW's successful flight testing validates the feasibility of compliant structures in real-world conditions, while the AdAR's telescopic design offers insights into managing structural complexity. However, both systems highlight the critical need for advanced materials with enhanced fatigue resistance, integrated sensing and control systems, and standardized testing and certification protocols. 

In addition, the commercial viability of morphing wing technology depends on balancing performance benefits against implementation costs and operational complexity. Our analysis indicates that partial morphing approaches, such as the MACW, offer a more immediate path to commercial implementation due to their localized nature and simpler certification requirements. Full-span morphing systems like the AdAR may find initial applications in specialized operations where their performance benefits outweigh complexity considerations. Beyond traditional commercial aviation, morphing wing technology shows promise in emerging applications such as unmanned aerial vehicles (UAVs) \cite{ling2019aerial}. \revision{The enhanced efficiency and adaptability could particularly benefit medical delivery drones and other specialized applications requiring operational flexibility.}

\section{Conclusion and Outlook}\label{sec:conclusion}

As commercial aviation faces increasing demands for fuel efficiency and operational flexibility, morphing wing technology offers promising solutions through adaptive aerodynamic surfaces. This paper has contributed a systematic classification framework for morphing wing designs and provided comprehensive analysis of two distinct implementation approaches. Through examination of historical developments and current case studies, we evaluated the Mission Adaptive Compliant Wing (MACW) and Adaptive Aspect Ratio (AdAR) systems, demonstrating performance improvements of up to 25\% in drag reduction and 40\% in control authority per degree deflection. Our results reveal that while partial morphing approaches offer more immediate implementation pathways, full-span morphing systems provide greater potential for overall flight optimization. These findings suggest that morphing wing technology is technically viable for commercial aviation, though certification and maintenance considerations remain critical challenges. Future developments in materials science and control systems, combined with standardized certification protocols, will be essential for widespread adoption in next-generation aircraft design.

\bibliographystyle{IEEEtran}
\bibliography{refs}  

\begin{thebibliography}{10}
\providecommand{\url}[1]{#1}
\csname url@rmstyle\endcsname
\providecommand{\newblock}{\relax}
\providecommand{\bibinfo}[2]{#2}
\providecommand\BIBentrySTDinterwordspacing{\spaceskip=0pt\relax}
\providecommand\BIBentryALTinterwordstretchfactor{4}
\providecommand\BIBentryALTinterwordspacing{\spaceskip=\fontdimen2\font plus
\BIBentryALTinterwordstretchfactor\fontdimen3\font minus \fontdimen4\font\relax}
\providecommand\BIBforeignlanguage[2]{{%
\expandafter\ifx\csname l@#1\endcsname\relax
\typeout{** WARNING: IEEEtran.bst: No hyphenation pattern has been}%
\typeout{** loaded for the language `#1'. Using the pattern for}%
\typeout{** the default language instead.}%
\else
\language=\csname l@#1\endcsname
\fi
#2}}

\bibitem{boeing2023cmo}
{Boeing Commercial Airplanes}, ``Commercial market outlook 2023-2042,'' \url{https://cmo.boeing.com}, 2023, accessed: 2024-04-19.

\bibitem{investors2023boeing}
{Boeing Investor Relations}, ``Boeing forecasts demand for 42,600 new commercial jets over next 20 years,'' \url{https://investors.boeing.com/investors/news/press-release-details/2023/Boeing-Forecasts-Demand-for-42600-New-Commercial-Jets-Over-Next-20-Years/default.aspx}, June 2023, accessed: 2024-04-19.

\bibitem{gorn2021high}
M.~H. Gorn and G.~De~Chiara, ``High-speed and high-altitude flight: {The} first {X}-planes,'' \emph{X-Planes from the X-1 to the X-60: An Illustrated History}, pp. 2--26, 2021.

\bibitem{choi2017idea}
C.~Choi, ``The 100-year-old idea that could change flight,'' \url{https://www.pbs.org/wgbh/nova/article/morphing-wings}, PBS NOVA, Feb. 2017, accessed: 2024-04-19.

\bibitem{frost2022global}
{Frost \& Sullivan}, ``Global commercial aircraft engine growth opportunities,'' \url{https://www.marketresearch.com/Frost-Sullivan-v383/Global-Commercial-Aircraft-Engine-Growth-31538714}, 2022, accessed: 2024-04-19.

\bibitem{boeing2015c40}
{Boeing Defense}, ``{C-40A} military transport aircraft,'' \url{https://www.boeing.com/defense/c-40a}, 2015, accessed: 2024-04-19.

\bibitem{milstein2007shape}
M.~Milstein, ``Shape shifters,'' \url{https://www.smithsonianmag.com/air-space-magazine/shape-shifters-15512296}, Air \& Space Magazine, 2007, accessed: 2024-04-19.

\bibitem{merlin2020green}
P.~W. Merlin, ``Green light for green flight: {NASA's} contributions to environmentally responsible aviation,'' National Aeronautics and Space Administration, Tech. Rep. NASA/SP-2020-4610, 2020.

\bibitem{ling2019aerial}
G.~Ling and N.~Draghic, ``Aerial drones for blood delivery,'' \emph{Transfusion}, vol.~59, no.~S2, pp. 1608--1611, 2019.

\bibitem{ninian2017design}
D.~Ninian and S.~M. Dakka, ``Design, development and testing of shape shifting wing model,'' \emph{Aerospace}, vol.~4, no.~4, p.~52, 2017.

\bibitem{nasm2022researching}
{National Air and Space Museum}, ``Researching the {Wright} way,'' \url{https://airandspace.si.edu/explore/stories/researching-wright-way}, 2022, accessed: 2024-04-19.

\bibitem{biolocomotion2011wright}
{Biolocomotion Lab}, ``The {Wright} flyer's warping wings,'' \url{https://blogs.bu.edu/biolocomotion/2011/10/18/the-wright-brothers-flyer}, Boston University, Oct. 2011, accessed: 2024-04-19.

\bibitem{nasa2023warping}
{NASA Glenn Research Center}, ``Wing warping interactive,'' \url{https://www1.grc.nasa.gov/beginners-guide-to-aeronautics/wing-warping-interactive}, June 2023, accessed: 2024-04-19.

\bibitem{crouch2008oldies}
T.~D. Crouch, ``Where do ailerons come from?'' \url{https://www.smithsonianmag.com/air-space-magazine/oldies-and-oddities-where-do-ailerons-come-from-40077712}, Smithsonian Magazine, 2008, accessed: 2024-04-19.

\bibitem{airforce2020f14}
{Air Force Technology}, ``{F-14} {Tomcat}: System overview and capabilities,'' \url{https://www.airforce-technology.com/projects/f14}, Dec. 2020, accessed: 2024-04-19.

\bibitem{nasa2023aeroelastic}
{NASA}, ``{F/A-18} active aeroelastic wing,'' \url{https://www.nasa.gov/reference/active-aeroelastic-wing}, Sept. 2023, accessed: 2024-04-19.

\bibitem{dibley2005development}
R.~P. Dibley, M.~J. Allen, R.~Clarke, J.~Gera, and J.~Hodgkinson, ``Development and testing of control laws for the active aeroelastic wing program,'' in \emph{AIAA Atmospheric Flight Mechanics Conference and Exhibit}, San Francisco, California, 2005, pp. AIAA--2005--6314.

\bibitem{barr2005wing}
L.~Barr, ``Wing warping could change shape of future aircraft,'' \url{https://www.af.mil/News/Article-Display/Article/134883/wing-warping-could-change-shape-of-future-aircraft}, U.S. Air Force, 2005, accessed: 2024-04-19.

\bibitem{woods2015adaptive}
B.~K.~S. Woods and M.~I. Friswell, ``The adaptive aspect ratio morphing wing: Design concept and low fidelity skin optimization,'' \emph{Aerospace Science and Technology}, vol.~42, pp. 209--217, 2015.

\bibitem{flexsys2024flexfoil}
{FlexSys Inc.}, ``{FlexFoil} compliant control surfaces,'' \url{https://www.flxsys.com/flexfoil}, 2024, accessed: 2024-04-19.

\bibitem{afshohteh2023sma}
A.~Afshohteh, ``Shape memory alloys,'' \url{https://www.linkedin.com/pulse/shape-memory-alloys-afshin-ashofteh-phd-mba-9szbf}, 2023, accessed: 2024-04-19.

\bibitem{friswell2006morphing}
M.~I. Friswell and D.~J. Inman, ``Morphing concepts for {UAVs},'' in \emph{21st Bristol UAV Systems Conference}, Bristol, UK, 2006, pp. 13.1--13.8.

\bibitem{kota2024future}
S.~Kota, ``Future airplanes will fly on twistable wings,'' \url{https://spectrum.ieee.org/future-airplanes-will-fly-on-twistable-wings}, IEEE Spectrum, Mar. 2024, accessed: 2024-04-19.

\bibitem{smith1992variable}
J.~W. Smith, ``Variable-camber systems integration and operational performance of the {AFTI/F-111} mission adaptive wing,'' National Aeronautics and Space Administration, Technical Paper NASA-TP-4370, 1992.

\bibitem{kota2009mission}
S.~Kota, R.~Osborn, G.~Ervin, D.~Maric, P.~Flick, and D.~Paul, ``Mission adaptive compliant wing: Design, fabrication and flight test,'' in \emph{RTO Applied Vehicle Technology Panel Symposium}, ser. MP-AVT-168, Evora, Portugal, 2009, pp. 1--18.

\end{thebibliography}

\end{document}